 \documentclass{article}

\usepackage{epsfig} 
\usepackage{amsfonts}
\usepackage{graphicx}

 \usepackage{amssymb}

\date{\today}

\newcommand{\insertplot}[5]{\begin{figure}
 \hfill\hbox to 0.05in{\vbox to #5in{\vfill
 \inputplot{#1}{#4}{#5}}\hfill}
 \hfill\vspace{-.1in}
 \caption{#2}\label{#3}
 \end{figure}}
 \newcommand{\inputplot}[3]{
 \special{ps: plotfile #1}
}
\newcounter{fig}

\newcommand{\ee}{\end{equation}}
\newcommand{\eea}{\end{eqnarray}}
\newcommand{\be}{\begin{equation}}
\newcommand{\bea}{\begin{eqnarray}}

\begin{document}

\title{ \bf 
Non-linear $Q$-clouds around Kerr black holes
}

\author{
{\large Carlos Herdeiro},  
{\large Eugen Radu}
and
{\large Helgi R\'unarsson}
\\ 
\\
{\small Departamento de F\'\i sica da Universidade de Aveiro and I3N} \\ 
{\small   Campus de Santiago, 3810-183 Aveiro, Portugal}
}
\date{September 2014}
\maketitle

\begin{abstract} 
\textit{$Q$-balls} are regular extended `objects' that exist for some non-gravitating, self-interacting, scalar field theories with a global, continuous, internal symmetry, on Minkowski spacetime. Here, analogous objects are also shown to exist around rotating (Kerr) black holes, as non-linear bound states of a test scalar field. We dub such configurations \textit{$Q$-clouds}. We focus on a complex massive scalar field with quartic plus hexic self-interactions. Without the self-interactions, linear clouds have been shown to exist, in synchronous rotation with the black hole horizon, along 1-dimensional subspaces  - \textit{existence lines} - of the Kerr 2-dimensional parameter space. They are \textit{zero modes} of the superradiant instability. Non-linear $Q$-clouds, on the other hand, are also in synchronous rotation with the black hole horizon; but they exist on a 2-dimensional subspace, delimited by a minimal horizon angular velocity and by an appropriate existence line, wherein the non-linear terms become irrelevant and the $Q$-cloud reduces to a linear cloud. Thus, $Q$-clouds provide an example of scalar bound states around Kerr black holes which, generically, are not zero modes of the superradiant instability. We describe some physical properties of $Q$-clouds, whose backreaction leads to a new family of hairy black holes, continuously connected to the Kerr family.  

\end{abstract}

\section{Introduction}
The discovery of Kerr black holes (BHs) with scalar hair~\cite{Herdeiro:2014goa}, continuously connected to the standard Kerr metric~\cite{Kerr:1963ud}, presents a qualitatively new example of asymptotically flat, regular on and outside an event horizon, hairy BHs~\cite{Herdeiro:2014ima,Herdeiro:2014jaa}. These solutions are anchored to the condition
\begin{equation}
\frac{w}{m}=\Omega_H , 
\label{sync}
\end{equation}
between the BH horizon angular velocity, $\Omega_H$, the scalar field frequency, $w$, and the scalar field azimuthal harmonic index, $m$. For Kerr BHs with scalar hair, two different possible interpretations of this condition are: 
\begin{description}
\item[(i)] it describes \textit{zero modes} (i.e. modes at the threshold) of superradiant instabilities, obtained by studying linear test fields on the Kerr background;
\item[(ii)] it describes the absence of scalar flux through the BH horizon, a necessary condition for the existence of scalar  (linear or non-linear)  \textit{bound states} on any rotating BH background.
\end{description}

Interpretation ${\bf (i)}$, on the one hand, comes about from considerations within \textit{linear theory}. A massive test scalar field mode of the form 
\begin{equation}
\Phi= e^{-iwt}e^{im\varphi}\phi(r,\theta) \ , 
\label{scalar-ansatz}
\end{equation}
on the Kerr BH background in standard Boyer-Lindquist (BL) coordinates, triggers a superradiant instability if $w<m\Omega_H$~\cite{Press:1972zz}. As such,  in~\cite{Herdeiro:2014goa} (see also~\cite{Dias:2011at}),  (\ref{sync}) was interpreted as describing the threshold of superradiant instabilities for this scalar field on the Kerr background. Then, the new family of Kerr BHs with scalar hair is naturally seen as branching off from the Kerr solution at the onset of a classical instability. This interpretation suggests that there is a \textit{general mechanism} connecting superradiant instabilities and new families of hairy BH solutions~\cite{Herdeiro:2014goa,Herdeiro:2014ima}: whenever a test field exhibits superradiant instabilities on a BH background, a new family of BH solutions with hair (of that field) should exist, continuously connecting to the original BH family on the subset of solutions that allow zero modes of the instability.

The zero modes of the instability are gravitationally trapped scalar field modes that neither grow, nor decay, computed in linear theory; thus ignoring their backreaction. As such they are bound states and have been called \textit{scalar clouds}~\cite{Hod:2012px,Hod:2013zza,Herdeiro:2014goa,Hod:2014baa,Benone:2014ssa}.\footnote{See \cite{Degollado:2013eqa,Sampaio:2014swa} for \textit{marginal} scalar and Proca clouds around charged BHs.}. These modes are found along 1-dimensional subspaces, called \textit{existence lines}, of the Kerr 2-dimensional parameter space. Since the Klein-Gordon equation separates on Kerr (in BL coordinates)~\cite{Brill:1972xj}, using an ansatz of the form $\Phi\sim e^{-iwt}e^{im\varphi}S_{lm}(\theta)R_{nlm}(r)$, where $S_{lm}(\theta)$ are the spheroidal harmonics and $n$ a node counting parameter for the radial function, $R_{nlm}(r)$, a given linear cloud is labelled by three `quantum numbers': $(n,l,m)$. Imposing condition (\ref{sync}), one finds that for each BH mass, the cloud is only possible for a specific value of $\Omega_H=w/m$. Alternatively, for a fixed Kerr background, only some clouds, and hence some frequencies are possible. This is effectively a quantization condition. These scalar clouds have a parallelism with the atomic orbitals of elementary quantum mechanics.

\bigskip

Interpretation ${\bf (ii)}$, on the other hand, comes from the fact that the null generator of the horizon of a stationary and axi-symmetric BH (not necessarily Kerr), $\chi=\partial_t+\Omega_H\partial_\varphi$, in coordinates adapted to the symmetries, preserves the scalar field (\ref{scalar-ansatz}), i.e. $k\Phi=0$, when condition (\ref{sync}) holds. Thus, this condition guarantees the absence of scalar flux through the horizon, a necessary requirement for an equilibrium state. This interpretation makes no reference to the phenomenon of superradiance and is valid beyond linear theory; it suggests that condition (\ref{sync}) may allow the existence of a broader set of hairy solutions than those associated to interpretation ${\bf (i)}$: solutions relying on  \textit{non-linear effects}.

\bigskip

In this paper we provide an example -- within the test field approximation -- of the latter type of solutions. We show that $Q$-balls, a type of scalar solitons known to exist around Minkowski spacetime~\cite{Coleman:1985ki,Lee:1991ax}, also exist as non-linear (test) bound states on the Kerr background, obeying condition (\ref{sync}). We call such solutions \textit{non-linear $Q$-clouds around Kerr BHs.}

\bigskip

$Q$-balls are complex scalar field solitons, obtained with a non-renormalizable self-interaction, arising in some effective field theories. 
These non-topological solitons circumvent the standard Derrick-type argument~\cite{Derrick:1964ww} by virtue of having a
time-dependent phase for the scalar field.
The global phase-invariance of the scalar field theory leads to a conserved Noether
charge $Q$, corresponding to particle number.
Such configurations have a rich structure and
found a variety
of physically interesting applications;
for example, they appear in supersymmetric generalizations of the standard model \cite{Kusenko:1997zq}, 
and have been suggested to generate baryon number or to be dark matter candidates~\cite{Kusenko:1997si}.

Here, we report that $Q$-balls can become $Q$-clouds, when replacing the near (Minkowski) origin region with a BH horizon\footnote{
Boson shells harbouring BHs have been studied in \cite{Kleihaus:2010ep}.
However, these solutions require a V-shaped scalar potential which
is not of the form (\ref{U}).}. This fits into a general pattern observed in soliton physics~\cite{Bizon:1994dh,Volkov:1998cc,Herdeiro:2014ima}; for the case discussed herein, however, the BH \textit{must rotate}, by virtue of condition (\ref{sync}), and the scalar field's frequency is not arbitrary. The relation between the scalar field's parameters and the horizon angular velocity can actually be seen as a \textit {rotation synchronization condition}, as discussed in \cite{Benone:2014ssa}. 

A distinctive new feature of the non-linear $Q$-clouds, when compared to the linear scalar clouds,
is that these solutions exist for a \textit{2-dimensional} region of the Kerr BHs parameter space, rather than just on 1-dimensional existence lines. For a specific $Q$-cloud, labelled by the integer $m$, this 2-dimensional space is bounded by a minimal angular velocity -- which shows these $Q$-clouds are supported by rotation -- and the existence line of linear clouds with the same $m$ and $n=0$, $l=m$. For fixed $m$, this particular existence line precisely separates Kerr BHs that are stable and unstable against superradiant instabilities triggered by this $m$-mode~\cite{Benone:2014ssa}; $Q$-clouds only exist in the stable region.  As the existence line that delimits their domain is approached, $Q$-clouds reduce to linear clouds. In this limit, the self-interaction terms (of the type we consider) become irrelevant for bound state solutions.

\bigskip

Finally, let us mention yet a different example, reported recently in~\cite{Brihaye:2014nba}, of how non-linear effects allow the existence of BHs with scalar hair, still anchored to condition (\ref{sync}). This example concerns Myers-Perry BHs~\cite{Myers:1986un} in five spacetime dimensions. In such backgrounds, a test massive scalar field does not admit bound state solutions (with real frequency), which may be regarded as a direct consequence of the incompatibility of the bound state and superradiant conditions~\cite{Cardoso:2005vk,Kunduri:2006qa}. Still, hairy BH solutions are found,  but which are  \textit{not continuously connected} to the Myers-Perry BHs in terms of the global charges (mass, angular momentum and Noether charge). In some limit, however, the horizon properties and local geometry of these hairy BHs becomes arbitrarily close to that of a particular sub-family of vacuum Myers-Perry BHs. As such they were dubbed in~\cite{Brihaye:2014nba} \textit{Myers-Perry BHs with scalar hair and a mass gap}. This gap, or discontinuity, in the global charges stands out as a signature that these hairy solutions rely on non-linear effects.

\section{The model}

We consider the action for a complex scalar field with self-interactions:
\begin{equation}
\label{action}
S=-\int \left[ 
   \frac{1}{2} g^{\mu\nu}\left( \Phi_{, \, \mu}^* \Phi_{, \, \nu} + \Phi _
{, \, \nu}^* \Phi _{, \, \mu} \right) + U( \left| \Phi \right|) 
 \right] \sqrt{-g} d^4x
\ . 
\end{equation}
The asterisk denotes complex conjugation 
and $U$ denotes the scalar potential; a usual choice in the $Q$-ball literature is 
\begin{equation}
U(|\Phi|) =  \mu^2 |\Phi|^2-\lambda |\Phi|^4 +\beta |\Phi|^6,
\label{U} 
\end{equation}
with $\mu$ being the boson mass and $\lambda,\beta>0$.
This potential is chosen such that non-topological soliton solutions
exist in a flat spacetime background, see $e.g.$ the discussion in \cite{Volkov:2002aj}.

Variation of (\ref{action}) with respect to the scalar field
leads to the non-linear Klein-Gordon (KG) equation,
\begin{equation}
\label{KG}
 \Box\Phi= \frac{\partial U}{\partial\left|\Phi\right|^2}\Phi \ ,
\end{equation}
where $\Box$ represents the covariant d'Alembert operator.
The stress-energy tensor $T_{\mu\nu}$ of the scalar field is
\begin{eqnarray}
T_{\mu \nu} 
=  \Phi_{, \, \mu}^*\Phi_{, \, \nu}
+\Phi_{, \, \nu}^*\Phi_{, \, \mu} -g_{\mu\nu} \left[ \frac{g^{\alpha\beta} }{2} 
\left( \Phi_{, \, \alpha}^*\Phi_{, \, \beta}+
\Phi_{, \, \beta}^*\Phi_{, \, \alpha} \right)+U(|\Phi|)\right]
 \ .
\label{tmunu} 
\end{eqnarray}

For the background metric,
we consider a general 
ansatz with two Killing vectors   $\xi=\partial_t$ and $\eta=\partial_\varphi$ (with $t$ and $\varphi$
the time and azimuthal coordinates, respectively),
which in an appropriate coordinate system
can be written as
\begin{eqnarray}
\label{metric-ansatz}
ds^2= g_{rr}dr^2+g_{\theta \theta} d\theta^2 +g_{\varphi\varphi}d\varphi^2+2 g_{\varphi t}d\varphi dt +g_{tt} dt^2 \ .
\end{eqnarray}
$g_{\mu\nu}$ 
are functions of the spherical coordinates $r$ and $\theta$ only. We assume asymptotic flatness. Thus, as $r\to \infty$, $g_{rr} \to 1$,
$g_{\theta \theta} \to r^2$,
$g_{\varphi\varphi} \to r^2\sin^2 \theta$,
$g_{\varphi t} \to 0$
and
$g_{tt} \to -1$.
We also assume the existence of an event horizon, located
at a constant value of $r=r_H$.
This 
is a Killing horizon of the Killing vector field
$\chi=\xi+\Omega_H \eta$,
where $\Omega_H$ is computed as
\begin{eqnarray}
\label{OmegaH}
\Omega_H=-\frac{\xi^2}{\xi \cdot \eta}\bigg |_{r_H}=-\frac{g_{tt}}{g_{t\varphi}}\bigg |_{r_H}.
\end{eqnarray}

The scalar field ansatz is of the form (\ref{scalar-ansatz}), 
 where $\phi$ is a real function, $w>0$ is the frequency and $m=\pm 1,\pm 2$\dots
is the azimuthal harmonic index. 
The fact
that the $(t, \varphi)$-dependences of $\Phi$  occur as phase factors only,
implies that $T_{\mu\nu}$ is $(t, \varphi)$-independent, which is required for
a configuration to be stationary and axisymmetric.
The energy-momentum tensor, however,  will 
 depend on both $m$ and $w$.

With the ansatz (\ref{scalar-ansatz}), (\ref{metric-ansatz}), 
the KG equation (\ref{KG})
reduces to
\begin{eqnarray}
\label{KG1}
\frac{1}{\sqrt{-g}}\frac{\partial}{\partial r}\left(g^{rr}\sqrt{-g}\frac{\partial \phi}{\partial r} \right)+
\frac{1}{\sqrt{-g}}\frac{\partial}{\partial \theta} \left(g^{\theta \theta}\sqrt{-g}\frac{\partial \phi}{\partial \theta} \right)
-\left(
m^2 g^{\varphi \varphi}-2g^{\varphi t} +w^2 g^{tt} 
\right)\phi
=(\mu^2-2 \lambda \phi^2+3\beta \phi^4)\phi.~{~}
\end{eqnarray}
We are interested in 
localized, particle-like solutions of this equation,
 with a finite scalar amplitude $\phi$ and a regular energy density distribution.
These axially symmetric configurations carry a nonzero 
  mass-energy and angular momentum,  which are defined as
\begin{eqnarray}
\label{scalar-charges}
E=-2\pi \int_{r_H}^\infty dr \int_0^\pi d\theta \sqrt{-g}T_t^t\ , \qquad 
J= 2\pi \int_{r_H}^\infty dr \int_0^\pi d\theta \sqrt{-g}T_\varphi^t \ .
 \end{eqnarray}
Moreover, a conserved charge $Q$ exists, associated with the complex scalar field $\Phi$, since the Lagrange density is invariant under the global phase transformation
$\Phi \to \Phi e^{i\alpha}$, 
 leading to the conserved current
 \begin{eqnarray}
\label{scalar-current}
j^{\mu}=-i\left[\Phi^* \partial^\mu\Phi+\Phi \partial^\mu\Phi^*\right]\ , \qquad \nabla_\mu j^{\mu}=0 \ .
 \end{eqnarray}
The corresponding conserved charge $Q$ is the integral of $j^t$ on spacelike slices.
One can easily see that in the absence of backreaction the following relation holds:
 \begin{eqnarray}
\label{rel1}
J=m Q\ ,
\end{eqnarray}
such that  angular momentum  is quantized.
Moreover,
in view of this relation, the spinning solutions
can be thought as corresponding to minima of energy with fixed angular
momentum.

In our approach, $Q$-clouds are found by solving the KG equation (\ref{KG1})
with suitable boundary conditions.  
Then the energy and angular momentum are computed
from the numerical output.
The boundary conditions result from the study of the solutions
on the boundary of the integration domain. 
The behavior of the scalar field as $r\to \infty$ must agree
with linear analysis: $\phi=f(\theta)  {e^{-\sqrt{\mu^2-w^2}r}}/{r}+\dots$;
thus $\phi|_{r=\infty}$=0, while the existence of a bound state requires $w <\mu $.
Also, axial symmetry and
regularity impose that the scalar field vanishes on the symmetry axis ($\theta=0,\pi$).
At the horizon, we suppose the existence of a power series expansion of the scalar field,
of the form
\begin{eqnarray}
\label{scalar-horizon}
\phi(r,\theta)=\phi_{0}( \theta)+\phi_{1}( \theta)(r-r_H)+\phi_{2}( \theta)(r-r_H)^2+\dots,
\end{eqnarray}
with finite coefficients $\phi_{k}$.
It turns out that, supposing $\phi_{0}\neq 0$, such an  expansion holds iff condition (\ref{sync}) holds. 
Thus, the synchronization (or no flux) condition is also required by regularity. 
After replacing (\ref{scalar-horizon})  into the KG equation, one obtains an involved condition between the coefficients
$\phi_{0}$, $\phi_{1}$ and $\phi_{2}$ 
which should be satisfied at $r=r_H$.
The explicit form of this condition depends on the coordinate system one chooses to work with,
and shall be discussed below.

\section{The solutions}

Similarly to previous works 
\cite{Volkov:2002aj,Kleihaus:2005me}, 
the  numerical solutions reported here have been found for 
the following parameters in 
the potential
(\ref{U}):
\begin{equation}
\lambda = 2,~~\beta=1,~~\mu^2=1.1~.
\label{param}
\end{equation} 
But solutions with other choices of $\lambda,\beta$
have also been considered.
In particular, preliminary results indicate that the constraints
on the potential $U$  required
for the existence of flat space $Q$-balls~\cite{Coleman:1985ki}, 
also hold for the BH background. 
Let us also remark that,
for given $(\mu,w,m)$,
solutions with other values of $\lambda$, $\beta$
can be generated by using the scaling symmetry
\begin{equation}
\phi^{(\lambda_2,\beta_2)}=\sqrt{\frac{\lambda_1}{\lambda_2}}\phi^{(\lambda_1,\beta_1)},~~
E^{(\lambda_2,\beta_2)}=\frac{\lambda_1}{\lambda_2}E^{(\lambda_1,\beta_1)},~~Q^{(\lambda_2,\beta_2)}=\frac{\lambda_1}{\lambda_2}Q^{(\lambda_1,\beta_1)},~~
{\rm with}~~\beta_2=\frac{\lambda_2^2}{\lambda_1}.
\label{symm}
\end{equation}

The KG (\ref{KG1}) equation has been solved by using a professional package, 
based on the iterative Newton-Raphson method \cite{schoen}.
The results found in this way for the known $Q$-balls case 
agree with those in the literature, 
obtained using other approaches.
Also, all quantities and variables of interest are expressed in natural units 
set by the scalar field mass $\mu$.

The $Q$-clouds discussed here are invariant under a reflection along  
the equatorial plane; these are commonly referred to, in this context, as \textit{even parity solutions}. Odd parity
solutions, on the other hand, should also exist, and their $Q$-ball limit has been considered in 
\cite{Volkov:2002aj,Kleihaus:2007vk}.
Also, we shall restrict our study to nodeless $Q$-balls, for which the scalar amplitude $\phi(r,\theta)$ has no nodes.

Finally, here we focus on a Kerr BH background, but similar solutions are likely to exist on any spinning BH.

\subsection{Limiting known cases: flat space $Q$-balls and Kerr linear clouds}

$Q$-clouds have two known limiting cases: (i) flat space $Q$-balls and (ii) Kerr linear clouds. We shall now briefly review some relevant properties of both these cases which are of interest to understand $Q$-clouds. 

\bigskip

Spinning $Q$-balls have been constructed  in 
\cite{Volkov:2002aj,Kleihaus:2005me};
a review of their properties can be found in \cite{Radu:2008pp}.
One imposes that the scalar field
vanishes at the origin, in a spherical coordinate system,  $\phi|_{r=0}=0$, which is implied by demanding regularity of the energy density.
Treating $w,m$ and the parameters in the potential $U$
as input variables, $Q$-balls 
exist only in a certain frequency range, 
$w_{min} < w < w_{max}=\mu$.
An estimate for $w_{min}$
is given by the condition 
\cite{Volkov:2002aj,Kleihaus:2005me}
\begin{eqnarray}
 w_{min}^2\sim  {\rm min} [U(\phi)/\phi^2]=\mu^2-\frac{\lambda^2}{4\beta}<w^2,
\end{eqnarray}
even though this limit could not be reached for spinning solutions.
At a critical value of the frequency in the interval $]w_{min}, w_{max}[$, 
both the $Q$-balls' mass-energy and angular momentum attain their minimum value,
from where they monotonically increase towards both limiting values
of the frequency.
Thus, considering the $Q$-ball's mass as a function of their Noether charge $Q$,
there are two branches of solutions, merging and ending
at the minimal charge and mass. 
$Q$-balls are stable along (most of) the lower frequency branch~\cite{Radu:2008pp},
where their mass is smaller than the mass of $Q$ free bosons.
Concerning their spatial distribution, for a given frequency, 
the amplitude $\phi(r,\theta)$, energy-momentum and charge
densities are maximal on the equatorial plane and the energy is concentrated in a toroidal
region encircling the symmetry-axis.

The limiting behaviour of the spinning $Q$-balls near the boundaries of the allowed frequency interval
is rather intricate, and has not yet been discussed in a systematic way  in the literature. 
It appears that 
both $E$ and $Q$
increase without bound at the limits of the $w$-interval, even though these limits are difficult to investigate.
Also, $Q$-balls become large there in terms of their spatial distribution;
as $w\to w_{min}$,
they can be viewed as squashed spheroids, homogeneously filled inside.
For $w\to w_{max}$, the solutions
also become large spheroids, but this time they are hollow, with the maximal energy density
concentrated at the surface and being close to zero everywhere else.  
Note that this
behaviour remains qualitatively the same for any $m>0$.

Finally, let us remark that flat spacetime $Q$-balls 
can be interpreted as \textit{scalarons} ($i.e.$ static solitons with $w=0$),
in a model with a shifted scalar field mass, for a new potential
$U = U_{(Q-ball) }- w^2|\Phi|^2$  \cite{Kleihaus:2013tba}.
Then the redefined potential $U$ is necessarily negative for some range of $|\Phi|$ which is realised by the solutions.
This interpretation, however, is lost for curved spacetime solutions.

\bigskip
  
\begin{figure}[h!]
\centering
\includegraphics[height=2.8in]{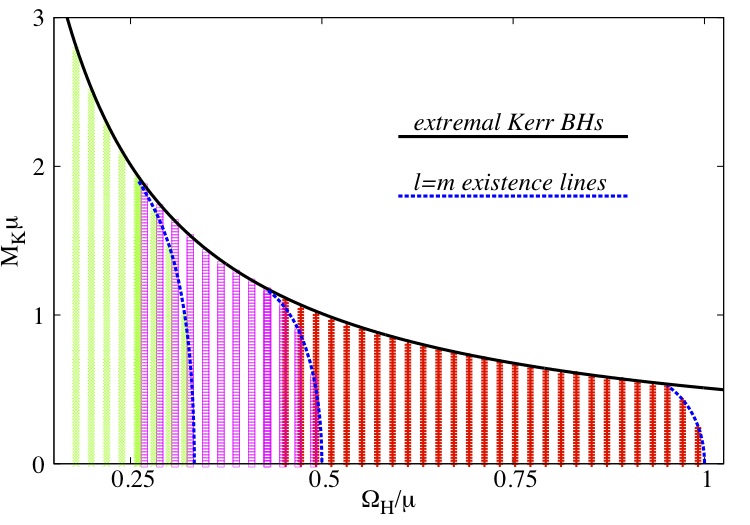}
\caption{Existence (blue dotted) lines with $n=0$, $m=l=1,2,3$, from right to left, respectively, for linear clouds on the Kerr background.  Kerr BHs exist below the solid black line, which corresponds to extremal Kerr solutions. For each $m$, $\Omega_H^{extremal}$ is the value of $\Omega_H$ at which the corresponding existence line intersects the curve of extremal BHs. The (vertical sets of) filling points in the diagram correspond to examples of $Q$-cloud solutions with $m=1$ (red/dark grey), $m=2$ (purple/medium grey) and $m=3$ (green/light grey). $Q$-clouds with a given value of $m$ exist between the existence line for linear clouds with $n=0$, $m=l$ and a minimal frequency.} 
\label{existencelines}
\end{figure}

A very different picture has been found in~\cite{Herdeiro:2014goa,Benone:2014ssa}
for the simpler case of a non-self-interacting
scalar field ($i.e.$ $\lambda=\beta=0$ in (\ref{U}))
on a fixed Kerr BH background.
These linear scalar clouds
satisfy the same boundary conditions as stated above.
However, their study is  simpler, since in this case
the KG equation (\ref{KG}) admits separation of variables, as described in the Introduction. 
Then the problem reduces to solving
an ordinary differential equation for 
the radial function $R_{nlm}(r)$.
This has been done in~\cite{Herdeiro:2014goa,Benone:2014ssa} and determines the position of the existence lines for a cloud with quantum numbers $(n,l,m)$. Three examples of these lines are exhibited in Fig. \ref{existencelines}, in a mass ($M$) vs. horizon angular velocity ($\Omega_H$) diagram for Kerr BHs.
Analytical estimates for these lines have been found in 
\cite{Hod:2012px,Hod:2013zza}
for the case of a (nearly-)extremal Kerr BH.

As already mentioned in the Introduction, the existence line with $n=0$ and $l=m$ divides the Kerr parameter space in two regions. To the left (right) of the line stand the Kerr backgrounds which are superradiantly stable (unstable) against scalar field perturbations with azimuthal harmonic index $m$.

\subsection{Non-linear $Q$-clouds on the Kerr black hole background}

When turning on the scalar field self-interactions in the potential (\ref{U}), the non-linearities prevent a separation of variables, similar to the one discussed above. 
For given $(w,m)$,
the scalar field $\phi$ is a superposition of
spheroidal harmonics, whose amplitudes, however, differ from $R_{nlm}(r)$.
Then following \cite{Volkov:2002aj}, one can write
\begin{eqnarray}
\phi(r,\theta)=\sum_{k=0}^\infty f_k(r)S_{m+2k,m} (\theta),
\end{eqnarray}
which results in an infinite set of ordinary differential
equations for $f_k(r)$.
In principle, this set can be truncated for some $k_{max}$
and then solved numerically.
In our approach, however, we have chosen to solve
 directly the partial differential equation (\ref{KG1}). But instead of using Boyer-Lindquist coordinates -- 
 which yield a complicated boundary condition at  $r=r_H$,  in terms of the scalar function and its first and second derivatives -- we have used quasi-isotropic coordinates for Kerr (see e.g.~\cite{Cook:2000vr}).  In parallel, a large set of solutions have been computed  by using a radially shifted version of the Boyer-Lindquist coordinate system, for which $r\to r-\frac{a^2}{r_H}$.
Both these coordinate systems yield a near horizon expansion for the scalar field, (\ref{scalar-horizon}), with the $\phi_1$ term absent, thus allowing us to impose a standard Neumann boundary condition there.

Our central result in this work is
that {\it all flat space Q-ball solutions can be generalized to Q-clouds on a Kerr BH background}.
The BH parameters, however, are not arbitrary, as implied by condition (\ref{sync}).
The solutions are found by starting with a flat spacetime
configuration with given $(w,m)$ and increasing the size of the BH
(as given $e.g.$ by the event horizon area)
via the parameter $r_H$.  
We have constructed in a systematic way solutions with $m=1,2,3$ (around 20000 solutions for each case). A subset of the solutions for each of these values of $m$ has been plotted in Fig.~\ref{existencelines}.  When varying the horizon size, there are two 
possible behaviours for the solutions with a given $w$: 
\begin{itemize}
\item[(i)]

For $m\Omega_H^{extremal}/\mu\leq w/\mu <1$,  $Q$-clouds start from flat spacetime $Q$-balls (i.e. with zero horizon size)
 and end on the existence line for linear clouds with $n=0$, $l=m$. The value of $\Omega_H^{extremal}$ depends on $m$; for instance,  $\Omega_H^{extremal}\simeq 0.95,0.43,0.26$  for $m=1,2,3$, respectively (see $e.g.$~\cite{Hod:2013zza}). As the existence line is approached,
the amplitude of the scalar field amplitude 
decreases to zero and 
the linear scalar clouds are recovered.
 
\item[(ii)]
For  $m\Omega_H^{min}\leq w <m\Omega_H^{extremal}$,
any Kerr BH with $\Omega_H=w/m$ is allowed as a background for a $Q$-cloud solution.
In particular, one finds scalar clouds also on extremal Kerr BHs,
which provide one boundary for the domain of existence. Again, the value of $\Omega_H^{min}$ depends on $m$ and they seem to coincide, in terms of $w_{min}$, with the corresponding value for $Q$-balls.

\end{itemize}
The bottom line is that $Q$-clouds exist between the $l=m$, $n=0$ existence line for linear clouds and a minimal frequency/horizon angular velocity -- Fig. \ref{existencelines}. A number of configurations with $m=4,5$
have also been found;
thus we expect their existence for any $m\geq 1$.

\begin{figure}[h!]
\centering
\includegraphics[height=2.2in]{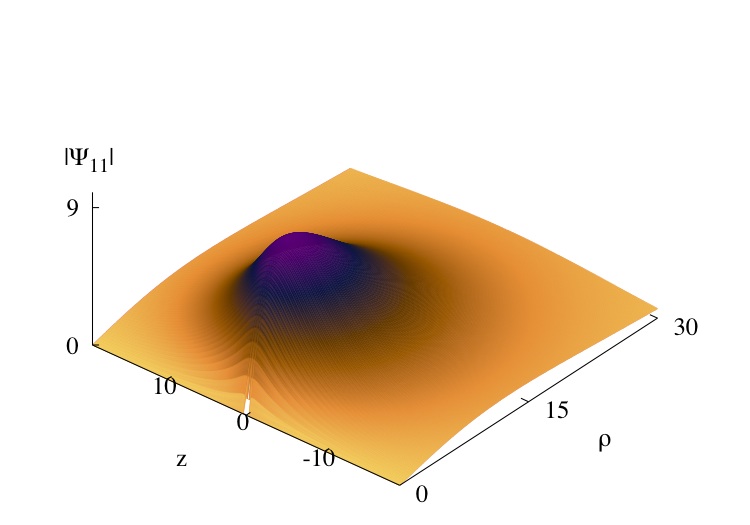}
\includegraphics[height=2.2in]{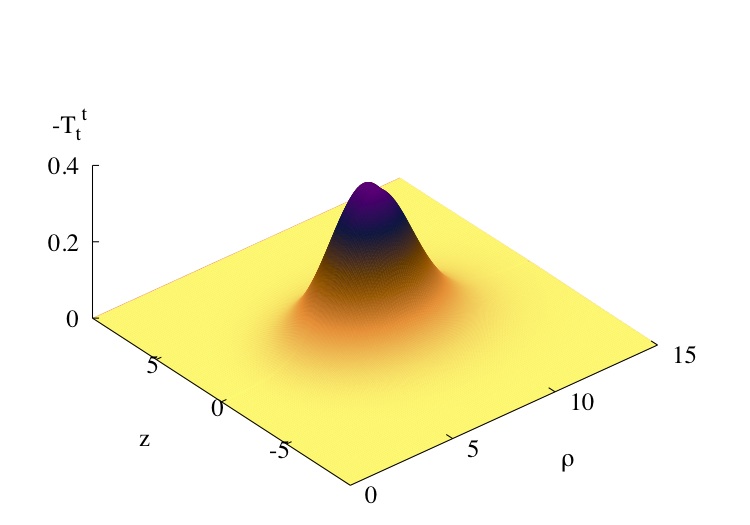}
\caption{The scalar field and the energy density are shown for a typical
 non-linear $Q$-cloud with $m=2$, $w =1$ (in terms of `polar' coordinates $\rho=r\sin \theta$,
 $z=r \cos \theta$).
The Kerr BH background has an event horizon radius (in quasi-isotropic coordinates) at
 $r_H=0.03$. 
} 
\label{densities}
\end{figure}

A generic $Q$-cloud solution has nonzero mass and angular momentum,
with a toroidal distribution for the corresponding densities -- Fig. \ref{densities}.
The scalar profile looks rather similar to that known in the flat spacetime limit
(with the region $r<r_H$ removed from it).
Note, however, that similarly to the behaviour observed for linear scalar clouds~\cite{Benone:2014ssa}, 
the scalar field amplitude does not vanish at the horizon,
approaching its maximum on the equatorial plane.
The energy density $-T_t^t$
vanishes on the symmetry axis, except for $m=1$ solutions.

\begin{figure}[h!]
\centering
\includegraphics[height=2.15in]{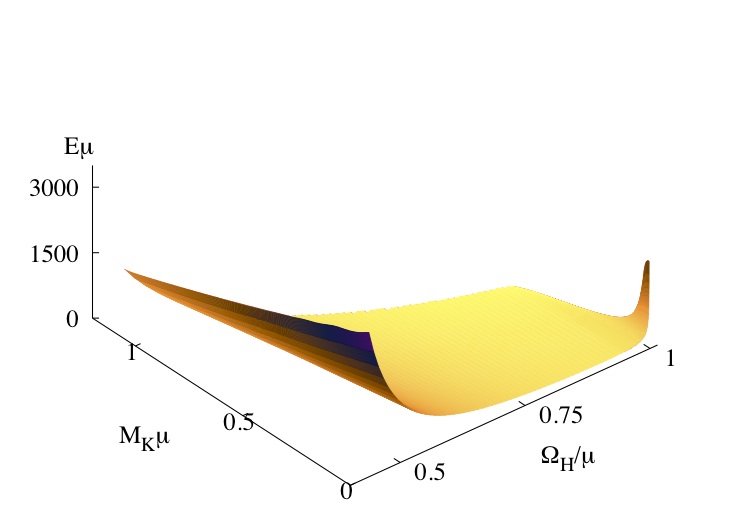}
\includegraphics[height=2.15in]{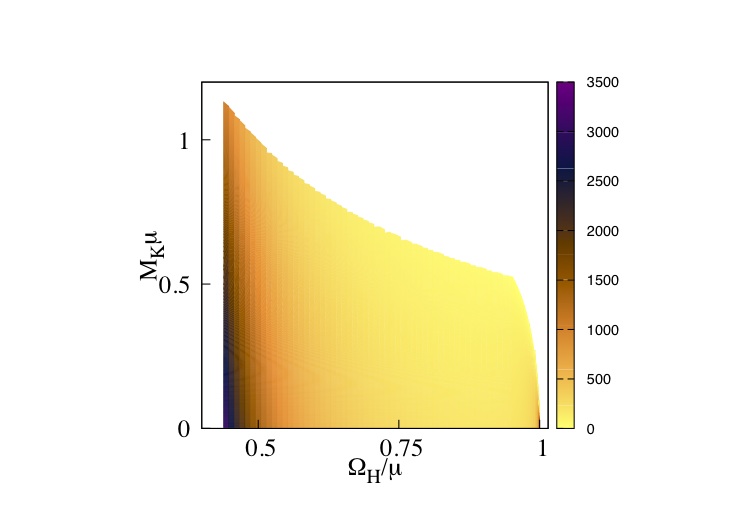}
\includegraphics[height=2.15in]{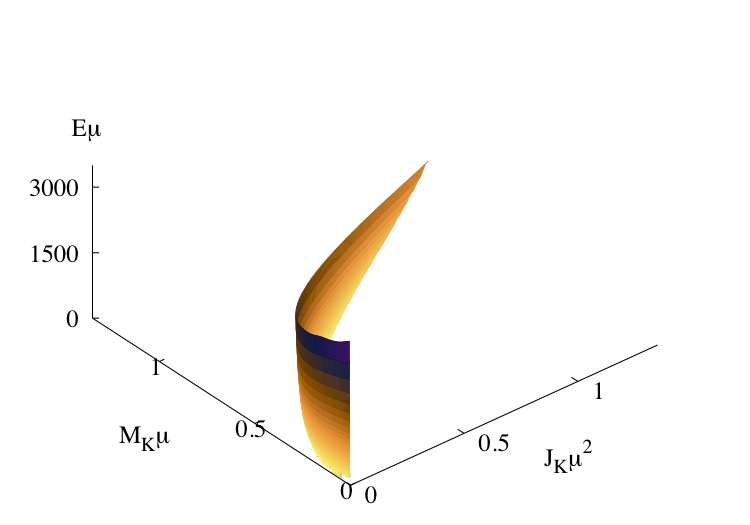}
\includegraphics[height=2.15in]{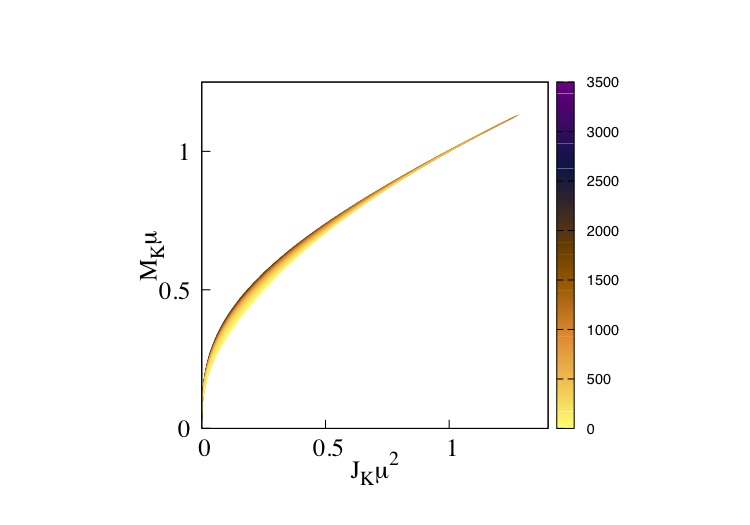}
\includegraphics[height=2.15in]{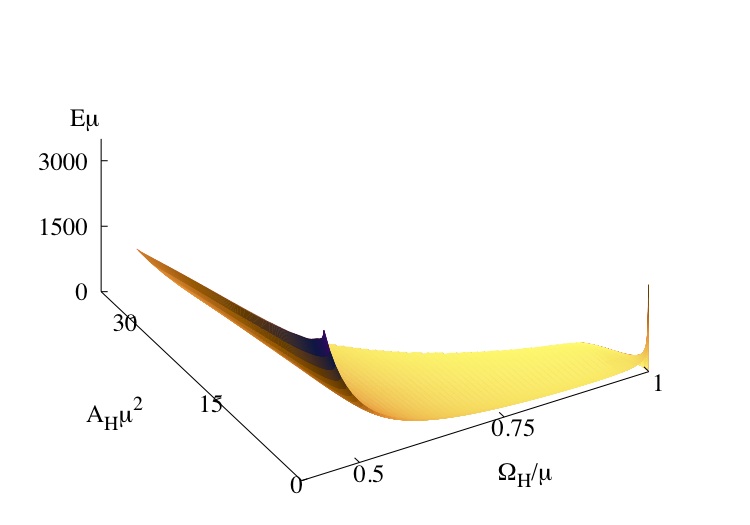}
\includegraphics[height=2.15in]{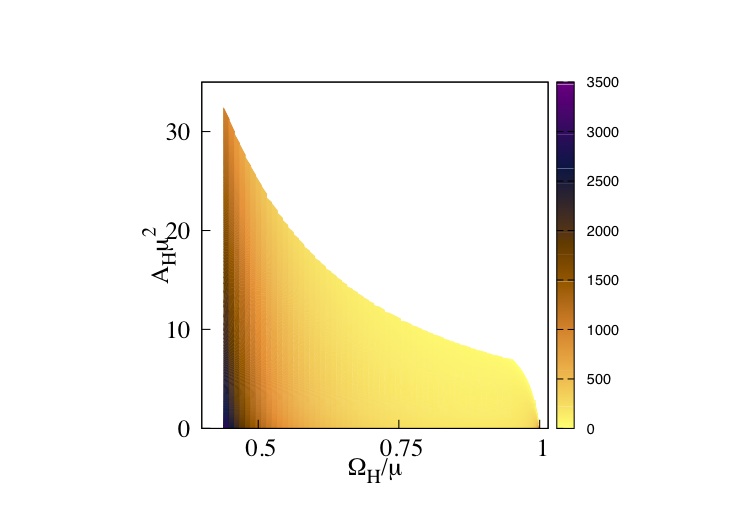}
\caption{$Q$-clouds energy, $E$, as a function three different combinations of parameters for the Kerr background, exhibited in both 3-dimensional (left panel) and 2-dimensional (right panel) plots: (top panel) mass $M_K$ and event horizon velocity $\Omega_H$; (middle panel) $M_K$ and angular momentum $J_K$; (lower panel) event horizon area $A_H$ and  $\Omega_H$.   
} 
\label{energy1}
\end{figure}

In Fig. \ref{energy1}, we plot the energy 
of $Q$-clouds as a function of several parameters of the Kerr BH background. These results have been obtained for $m=1$, but a similar picture has been found for $m=2,3$. A similar pattern is observed for the angular momentum $J$.

\newpage

\begin{figure}[h!]
\centering
\includegraphics[height=2.8in]{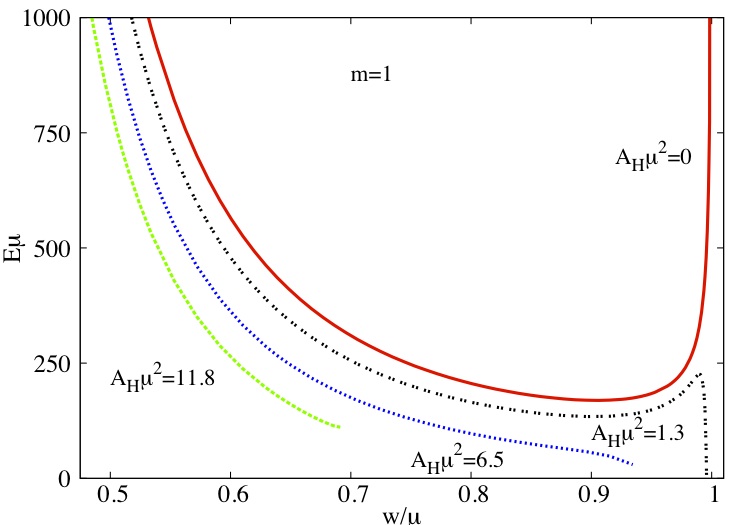}
\caption{The spectrum of $Q$-clouds for several
fixed values of the event horizon area $A_H$.
The solutions with $A_H=0$
are the flat space $Q$-balls.} 
\label{spectrum}
\end{figure}

Similarly to the flat spacetime case,
  the study of the solutions with $w\to w_{min}$
  is rather difficult, since
 both the energy and angular momentum of the $Q$-clouds 
 take very large values in this limit. 
 At the same time, and in contrast to flat space $Q$-balls, the charges of $Q$-clouds
 remain finite as the maximal frequency is approached. In this case, the BH background plays the role of a regulator.
 These limiting behaviours are illustrated in Fig. \ref{spectrum}, where we plot the energy spectrum of $m=1$ $Q$-balls,
 $E(w)$,
 for several fixed values of the horizon area $A_H$.
 One can see that $Q$-clouds on a `small'  Kerr BH
 approach  for $w_{max}<\mu$,
 a critical configuration with zero global charges; this configuration sits 
 on the corresponding ($n=0,~l=m$)  existence line.
On the other hand, $Q$-clouds on a `large' Kerr BH 
 end at a critical configuration with nonzero mass-energy and angular momentum;
 the corresponding Kerr backgrounds have $T_H=0$.

\section{Further remarks} 

We have shown that
the well-known flat space $Q$-balls 
possess generalizations on a rotating BH background -- $Q$-clouds. These bound states are 
in synchronous rotation with the BH horizon, i.e. they obey condition (\ref{sync}).
Remarkably,
the self-interactions allow the existence of non-linear clouds 
in a \textit{2-dimensional} subspace of the full parameter space of Kerr BHs.
This subspace is bounded by the flat spacetime $Q$-balls,
the  $n=0,l=m$ existence line of linear clouds~\cite{Benone:2014ssa}
and a critical curve delimited by the minimal frequency of the scalar field.
 
The backreaction of $Q$-clouds leads to a new family of Kerr BHs with scalar hair in the full Einstein-scalar field system, when this type of self interactions are included. The new solutions have quantitative and qualitative differences, relatively to the Kerr black holes with scalar hair found in~\cite{Herdeiro:2014goa}. We have obtained some examples of these solutions, which will be reported in detail somewhere else.

Finally, let us mention that we have found no bound state solutions on the Kerr background for self-interacting scalar fields with the renormalizable potential $U(|\Phi|) =  \mu^2 |\Phi|^2+\lambda |\Phi|^4$. This does not exclude, however, that such theories can support hairy BHs, under condition (\ref{sync}). In a similar spirit,  boson stars can exist for this potential~\cite{Colpi:1986ye,Schunck:2003kk},  but they trivialize in the flat space limit, since the potential does not support $Q$-balls.

\vspace{0.5cm} 
\noindent
{\bf\large Acknowledgements}\\ 
C.H and E.R. gratefully acknowledge support from the FCT-IF programme. 
 H.R. is funded by the FCT grant SFRH/BI/52523/2014. 
The work in this paper is also supported by the grants PTDC/FIS/116625/2010 and  NRHEP--295189-FP7-PEOPLE-2011-IRSES.

 \bibliographystyle{h-physrev4}
\bibliography{hbhs}

\end{document}